\documentclass[12pt,preprint]{aastex}

\shorttitle{RADIO EMISSION FROM LP 349-25}
\shortauthors{Phan-Bao et al.}

\begin{document}

\title{DISCOVERY OF RADIO EMISSION FROM THE TIGHT M8 BINARY: LP~349-25}

\author{NGOC PHAN-BAO,\altaffilmark{1} RACHEL A. OSTEN,\altaffilmark{2,3}
JEREMY LIM,\altaffilmark{4} EDUARDO L. MART\'IN,\altaffilmark{5,1} PAUL T.P. HO\altaffilmark{4,6}}

\altaffiltext{1}{University of Central Florida, Dept. of Physics, 
PO Box 162385, Orlando, FL 32816-2385; pngoc@physics.ucf.edu.}
\altaffiltext{2}{Astronomy Department, University of Maryland, College Park, MD 20742;
rosten@milkyway.gsfc.nasa.gov.}
\altaffiltext{3}{Hubble Fellow}
\altaffiltext{4}{Institute of Astronomy and Astrophysics, Academia Sinica,
P.O. Box 23-141, Taipei 106, Taiwan, R.O.C.; jlim@asiaa.sinica.edu.tw.}
\altaffiltext{5}{Instituto de Astrof\'{\i}sica de Canarias, 
C/ V\'{\i}a L\'actea s/n, E-38200 La Laguna (Tenerife), Spain; ege@iac.es.}
\altaffiltext{6}{Harvard-Smithsonian Center for Astrophysics, 60 Garden Street, Cambridge, MA 02138;
ho@cfa.harvard.edu.}

\begin{abstract}
We present radio observations of 8 ultracool dwarfs with
a narrow spectral type range (M8-M9.5) using the Very Large Array at 8.5~GHz. Only the tight M8 binary
LP~349-25 was detected. LP~349-25 is the tenth ultracool
dwarf system detected in radio and its trigonometric parallax $\pi = 67.6$~mas,
recently measured by Gatewood et al., makes it the furthest ultracool system 
detected by the Very Large Array to date, and the most radio-luminous outside
of obvious flaring activity or variability. With a separation of only 1.8~AU,
masses of the components of LP~349-25 can be measured precisely without any theoretical
assumptions (Forveille et al.), allowing us to clarify their fully-convective status and hence 
the kind of magnetic dynamo in these components which may play an important role to explain
our detection of radio emission from these objects. This also makes LP~349-25 an excellent target  
for further studies with better constraints on the correlations between X-ray, radio emission
and stellar parameters such as mass, age, temperature, and luminosity in ultracool dwarfs.
\end{abstract}


\section{INTRODUCTION}

Main sequence stars are expected to be fully convective if their mass lies below
a certain value, about 0.3--0.4~$M_{\odot}$ (M3$-$M4 spectral types), as suggested
by standard models; this probably shifts toward lower masses (0.1--0.2~$M_{\odot}$) due to the influence
of the magnetic field \citep{mullan}. In fully convective stars, a turbulent dynamo
that differs from the shell dynamo at work in partly convective Sun-like stars has been
proposed to take place \citep{durney, dobler, chabrier}. 
However, some recent observations
\citep{donati} have not supported such theoretical models. 
On one side, the model of Chabrier \& K\"uker (2006) predicts: (1) weak 
surface differential rotation, in agreement with the observations of Donati et al.; 
(2) non-axisymmetric large scale fields, in contradiction with these
observations.
On the other side, the model of Dobler et al. (2006) concludes: (1) large
scale axisymmetric fields, in agreement with the observations; (2) significant
surface differential rotation, in contradiction with the
observations.
Therefore understanding
how the magnetic field is generated in the fully convective stars and what the field properties
are is clearly important.

The atmospheres of such ultracool, very low mass stars are predominantly cool,
dense, and highly neutral, apparently precluding the build-up of magnetic stresses
thought to provide magnetic heating.
Yet, detections of observational signatures commonly associated with magnetic activity
in Sun-like and fully convective stars \citep{liebert, berger01,stelzer06}
suggest that the production mechanism is the same or nearly the same in very low mass
stars as for low mass and Sun-like stars.
Further study of the X-ray emissions \citep{neuhauser, stelzer06} and $H\alpha$ emissions \citep{gizis00, mohanty03} 
is one way to delimit the properties of magnetic heating in very low mass stars and indirectly
investigate the influence of the magnetic field.
Recent detections and studies of radio emission from ultracool dwarfs 
\citep{berger01, krish, berger02, burgasser05, blank, berger05, audard, osten06a, osten06b, berger06, hallinan} 
also offer us another important 
approach to understand the properties of magnetic fields in the fully convective stars,
as the magnetic field is likely intimately involved in the production mechanism of the
radio radiation from these objects.

In this paper, we present our radio observation of a sample of 8 ultracool dwarfs and
the detection of the M8 binary LP~349-25 
using the Very Large Array.
In \S~2 we present the observation and the data reduction, the properties of the radio
emission from LP~349-25 are given in \S~3, 
we discuss radio emission from LP~349-25 particularly and its radio mechanism in \S~4.

\section{SAMPLE, OBSERVATION AND DATA REDUCTION}

\subsection{Sample}

We selected a sample of 
nearby ultracool field dwarfs with spectral types ranging
over a narrow band of M8--L0 since 5 of 9 ultracool dwarfs detected at radio wavelengths
are M8--M9.5 dwarfs, giving a significant fraction of $\sim$56\%
(see Table~1 in \citealt{berger06} and references therein).
The rotational velocities of the stars
range from 4 to 35~km~s$^{-1}$, and the $H\alpha$ equivalent width ranges from 0 to 21~\AA.
Some known binary systems are also added, including the new tight M8 binary LP~349-25 
whose binary status was recently reported by \citet{forveille05}. 

In this paper, we present our radio observation results of 8 of these 
dwarfs. Their properties are listed in Table~\ref{sample}. 

\subsection{Observations and data reduction}
\begin{figure}
\centerline{\includegraphics[width=5in,angle=0]{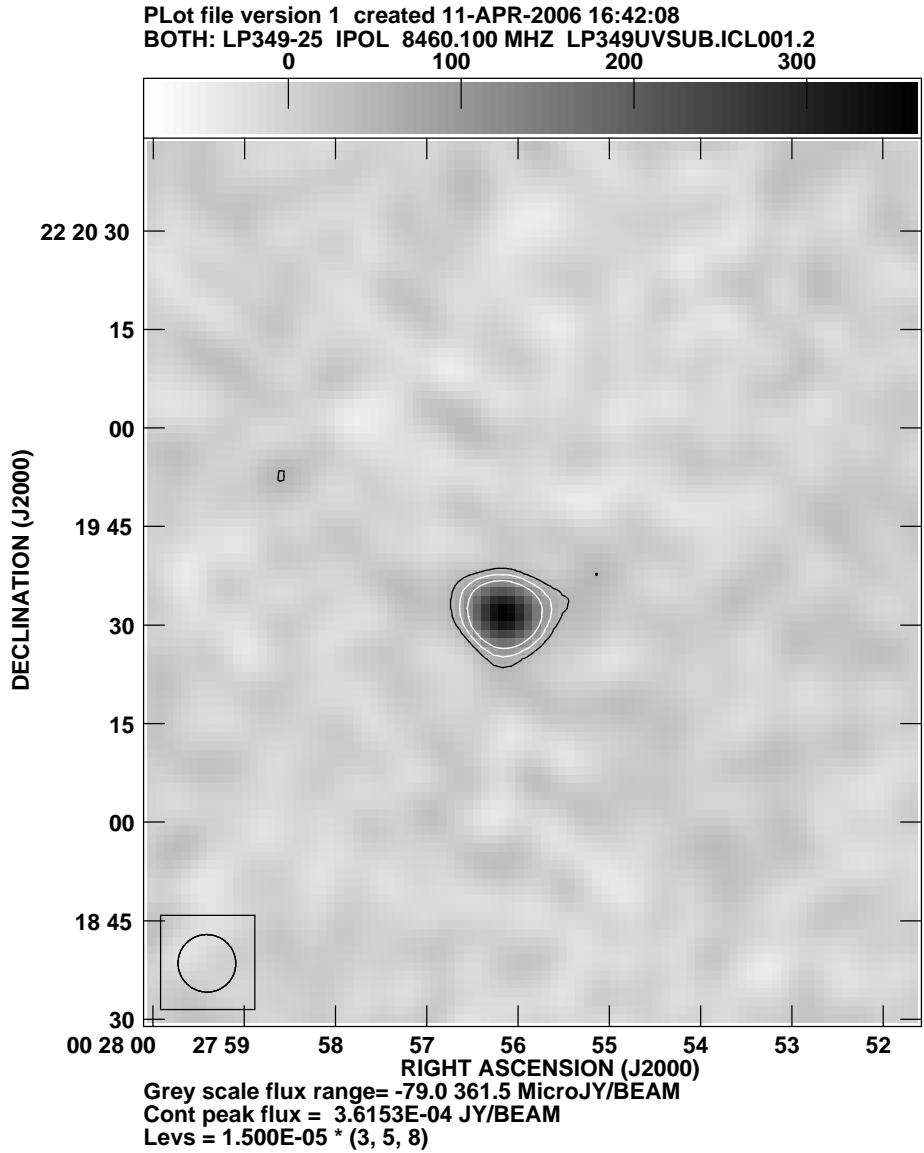}}
\caption{A VLA map of the LP~349-25 field at 8.5~GHz. The grey scale image is the map of total intensity.
Contours of total intensity are overplotted as multiples of the 1 $\sigma$ flux uncertainty; the 
3, 5 and 8 $\sigma$ contour levels are shown.
The beam size is also shown at the bottom left of the map.
\label{VLAimage}}
\end{figure}

We observed the 8 dwarfs from 2005 December 22 to 2006 January 02 with
the NRAO Very Large Array (VLA)\footnote{
The NRAO is a facility of the National Science Foundation operated under
cooperative agreement by Associated Universities, Inc.} 
when the array was the D configuration, except the observation of 2M~0140$+$27
in DnC, at 8.5~GHz (3.6~cm) using the standard continuum mode with 
2~$\times$~50~MHz contiguous bands. 
The flux density calibrators were 3C~48, 3C~147,
and 3C~286 and the
phase calibrators were selected to be within $10^{\circ}$ of the targets. 
The data was reduced and analyzed with the
Astronomical Image Processing System (AIPS).
Only LP~349-25 was detected at the observed frequency
of 8.5~GHz (Fig.~\ref{VLAimage}). 
In order to examine the emission from this object, we subtracted the visibilities 
of other radio sources in the field, and re-imaged the visibility dataset.  We extracted
the position and flux of the source in total intensity, and made an image of the dataset in
circular polarization (Stokes V).  In addition, we investigated the time variation of the source
using the task DFTPL within AIPS.

\subsection{Analysis}
Table~\ref{sample} gives the flux 
and upper limits for the detections/non-detections, respectively.
The position of the radio source near the center in the map of the field around LP~349-25
coincides with the expected position of LP~349-25 at the epoch of our observations (2005.967)
with its proper motion included 
($\mu$RA$=0.408$~arcsec~yr$^{-1}$, $\mu$DE$=-0.174$~arcsec~yr$^{-1}$, \citealt{lepine}).
We therefore conclude that the detected radio emission is from LP~349-25.
Our initial examination of maps made in total intensity (Stokes I) and circular polarization 
(Stokes V) show the existence of a weak source in the Stokes V map near this 
same position.  Further investigation,
however, reveals that the location of the source in the Stokes V map is offset with respect to
the position of the source in the Stokes I map by $\approx$ 10 arcseconds.
The right (RCP) and left (LCP) circularly polarized primary beams of the VLA are separated by
about 6\% of the half-width at half maximum of the antenna beam\footnote{http://www.vla.nrao.edu/astro/guides/vlas/current/node34.html}.  
This gives rise to a phenomenon known as ``beamsquint'' \citep{cotton}, and for observations at 8.5 GHz
the magnitude of the offset is 10.3 arcsec.    This makes the detection of circular
polarization intrinsic to the source suspect.
Finally, re-imaging of this field in Stokes V after subtracting the visibilities in total intensity
produces a lower significance ($<$3$\sigma$) of the source.  Based on the fluctuations in the
Stokes V image, we deduce a 3$\sigma$ upper limit of 48 $\mu$Jy, or a circular polarization
percentage $<$13\%.

The total extent of the observations of LP~349-25 was only $\sim$1.7 hours, precluding
the possibility of searching for rotational modulations, as this would correspond to
one rotation period only for very rapidly rotating objects \citep[e.g. TVLM~513-46546;][]{osten06a,hallinan}.
We extracted light curves of total intensity 
from the visibility dataset using time bins of 300 and 60 seconds
to search for short-time scale transient brightenings.
No obvious evidence of flare variability presented itself on visual inspection of either light curve.
Using the one-sided Kolmogorov-Smirnov statistic, we tried to see if we could reject the null
hypothesis that the fluxes are distributed uniformly in time and thereby establish evidence for
variability.  We compared the cumulative distribution function of the fluxes to that of a uniform
distribution in time, and conclude that the data are consistent with being drawn from a
uniform distribution of events with KS statistics of 0.05 and 0.01 for the 300 and 60 second
time binnings, respectively.  The KS probabilites are unity in both cases, 
thus we find no evidence for variability. 

\section{PROPERTIES OF THE RADIO EMISSION FROM LP 349-25}

The flux detected from LP~349-25 together with the parallax measurement of 67.6~mas
\citep{gatewood} imply a radio luminosity of 9.6$\times$10$^{13}$~erg~s$^{-1}$~Hz$^{-1}$.
This is $\approx$~3 times larger than the steady luminosities of previously detected
single very low mass dwarfs with similar spectral types (M8--M9). The makeup of the binary system
in LP~349-25 is likely M7.5V+M8.5V or M8V+M9V \citep[according to][]{forveille05},
and thus if the two components
are equal contributors to the radio flux the radio luminosity of each dwarf would be 
comparable to or slightly brighter than the steady flux levels
of the brightest apparently single dwarf previously detected.
The projected separation of the binary is
0.12\arcsec~or 1.78~AU at the distance of 14.8~pc, derived from
its trigonometric parallax measured recently \citep{gatewood}. 
The beam size of the image from 
our D array observations is 9.2\arcsec$\times$8.0\arcsec (major axis $\times$ minor
axis), with a position angle of 76$^{\circ}$, thus we cannot spatially resolve the two components.

With only one observed frequency of 8.5~GHz, we cannot constrain an emission mechanism. 
Previous papers have discussed two mechanisms: 
gyrosynchrotron emission from a population of mildly relativistic accelerated particles 
\citep{berger02, berger06, burgasser05, osten06a}, and
cyclotron maser emission \citep{hallinan}.
The brightness temperature of the emission is fundamental to constraining the emission
mechanism, but with two dwarfs both being potential sources of the radiation, there are
two unknowns in converting flux to brightness temperature:  The fraction of the total flux 
which each component contributes, and the size of the emitting region (being possibly different
for each dwarf).  We consider two
simplifying cases: (1) each dwarf contributes half the observed flux, with characteristic length
scale the radius of the dwarf (where the radius is $\approx$0.1~R$_{\odot}$) as suggested
by the previous observations \citep{leto, berger06}; (2) one dwarf contributes all of the emission, 
with length scale one tenth the dwarf radius as observed in a few M dwarfs \citep{lang, bastian}.  
Application of the relationship between brightness
temperature, frequency, flux density, and source size \citep{leto}, \\
\begin{equation}
T_{b} =\frac{4.33\times10^{5}}{\nu_{\rm GHz}^{2}} \frac{S_{\mu \rm Jy}}{\theta_{\rm mas}^{2}} {\rm K}
\end{equation}
where $S_{\mu \rm Jy}$ is the source flux in $\mu$Jy, $\nu_{\rm GHz}$ is the observing frequency in 
GHz, and $\theta_{\rm mas}$ is the angular size of the source in mas, yields for the
first case $T_{b}=1.1\times10^{9}$~K, and in the second case $T_{b}=2.2\times10^{11}$~K.
The high brightness temperatures indicate that the emission is nonthermal and could be consistent 
with gyrosynchrotron  emission \citep{dulk85}. One should note that larger size scales reduce the
estimated brightness temperatures.
In the second case, with $T_{b} \sim 10^{11}$~K a coherent mechanism, such as
cyclotron maser emission is also applicable.
Such a mechanism has been suggested by \citet{hallinan}
to account for the periodic variation of flux density and circular polarization in two other dwarfs.
\citet{hallinan} noted that the unmodulated flux of their target, TVLM~513-46546, could be attributable
to gyrosynchrotron emission with undetected circular polarization.  Clearly, more observations of
LP~349-25 are needed to determine its characteristics better.

\section{DISCUSSION}

Our sample of 8 late-M dwarfs with spectral types ranging from M8.0 to M9.5 produced only
one detection, that of a binary consistent with two M8 dwarfs. 
Including the objects in our survey with previous reports on late M-T dwarfs
\citep{krish, berger01, berger02, berger06, burgasser05, audard, osten06b},
the fraction of late-M and brown dwarfs detected at radio wavelengths is $\sim$10\% so far (see Table~1
in \citealt{berger06} and references therein, \citealt{osten06b}).
Interestingly, over a narrower range of spectral type, $\sim$M8$\le SpT\le $M9.5 a larger fraction, 18\%, 
of objects have been detected at radio wavelengths, including our sample. 

\citet{berger02} suggested that rotation may play a role in influencing which objects are detected
at radio wavelengths, by noting that the few objects with radio detections were among those
with the largest values of $v\sin i$.  The relationship between
rotation and the existence of large-scale magnetic fields, which may be necessary to
influence the production of radio emision, has been suggested for fully convective
stars by \citet{phan-bao}, based on measurements of longitudinal magnetic field strengths
in the active dM3.5e star EV~Lac ($v\sin i = 4.5$~km~s$^{-1}$) and absence in
the slower rotator dM5.5 HH~And ($v\sin i<1.2$~km~s$^{-1}$).
The fact that none of the fast rotators in our sample ($7 \leq v\sin i \leq 12$~km~s$^{-1}$)
provided radio detections
suggests that $v\sin i$ is not the dominant factor controlling the
production of detectable levels of radio emission.
Only the binary LP~349-25 was detected but unfortunately
the rotational velocities of its components have not been measured so far.
Thus, the correlation between radio emission and rotation is still unclear.

The ambiguous relationship between rotation and radio emission is strengthened upon
considering
the field M stars VB~8 and LHS~3003 (both spectral type M8), which have similar 
rotational velocities (8~km~s$^{-1}$ for VB~8 and 9~km~s$^{-1}$ for LHS~3003),
and are located at $d\sim$~6.5~pc.  
Despite the similarities in spectral type, $v\sin i$, and distance (which produces similar
sensivities), only LHS~3003 has shown radio quiescent emission \citep{burgasser05, krish}.
One possible explanation is that they were observed at different radio frequencies: \citet{krish}
observed VB~8 at 8.5~GHz and \citet{burgasser05} observed LHS~3003 at frequencies
of 4.8 and 8.6~GHz but they only detected the star at 4.8~GHz. 
This might suggest that the spectral peak of radio emission from ultracool and
brown dwarfs is closer to 4.8~GHz rather than 8.5~GHz, as discussed in
\citet{osten06a}.
Another possible explanation lies in the bias that inclination can introduce
in estimating rotation rates from $v\sin i$.  Photometric periods combined with
spectroscopic rotation velocities have the potential to nail down this discrepancy \citep[e.g.][]{bailer-jones}.

There is no explicit dependence between either emission mechanism proposed so far for radio
emission from very low mass dwarfs and rotation.  For two cases where radio emission has been
detected and rotational modulation of the emission has been determined, \citet{hallinan}
proposed a model whereby 
a large scale dipole or multipole with magnetic field
strength of a few kG at the stellar surface can produce cyclotron maser emission in regions
where the strong-field condition $\nu_{p}/\nu_{B}\le$ 1 is satisfied, where $\nu_{p}$ is the 
plasma frequency ($\approx 9000 \sqrt{n_{e}}$~Hz) and $\nu_{B}$ is the gyro-frequency
($=2.8 \times 10^{6}B$(G)~Hz); i.e. in regions of low electron density and high magnetic field
strength.
The recent detection of a strong
and large-scale axisymmetric magnetic field on a rapidly rotating M4 dwarf (V374~Peg, $v\sin i=36$~km~s$^{-1}$)
bolsters the suggestion of a link between large-scale fields, rapid rotation, and detectable levels
of varying radio emission,
but provides only a post hoc explanation for why some objects are detected at radio wavelengths.
Clearly, a detailed investigation of those objects which do show evidence of radio emission is needed,
along with suitable samples of objects with similar properties but lacking in radio detections, to tease
out any controlling parameters.

So far, there are no clear correlations between radio emission from ultracool and brown dwarfs
and their properties such as rotation, spectral types, and chromospheric activity; and radio emission
mechanisms are possibly gyrosynchrotron or coherent electron cyclotron maser emission.

\section{CONLUSION}

We present our radio observations of a sample of 8 late-M dwarfs, with spectral types
ranging over a narrow band of M8.0$-$M9.5. Only one of them has been detected, the M8 binary
LP~349-35. We have observed LP~349-25 at only 8.5~GHz, thus incoherent gyrosynchrotron
or coherent electron cyclotron maser emission could be applicable.
Observation of multiple radio frequencies will clarify the radio emission mechanism of the binary, and 
measurement of $v\sin i$ of each component will also be required to test whether any 
correlation with rotation is implied by this radio detection.  The lack of detections of
the fast rotators in our sample suggest a murky picture of the relationship between
radio emission and rotation (as measured by $v\sin i$).
Finally, with a separation of only 1.8~AU, derived from an angular separation of 0.12\arcsec~and $\pi = 67.6$~mas
\citep{forveille05, gatewood},
dynamical masses of each component of LP~349-25 are under measurement (T. Forveille, private communication).
As the masses of the components of LP~349-25 can be measured precisely without any theoretical assumptions, this
will therefore allow us to clarify their fully-convective status 
(e.g., 0.1--0.2~$M_{\odot}$, \citealt{mullan}), and hence the kind of magnetic dynamo in these
ultracool dwarf components which may play an important role to explain our detection of radio emission 
from these objects.
This also makes the system an excellent target for further studies with 
better constraints on the correlations between X-ray, 
radio emission and stellar parameters such as mass, age, temperature, and luminosity in ultracool dwarfs.

\acknowledgments
This paper presents the result of VLA project AP495. 
N. P.-B. and E. L. M. acknowledge financial support from NSF grant AST 0440520.
N. P.-B. thanks Y.-N. Su for helps to prepare observation scripts and P.-G. Gu for
useful discussion.
Support for this work was provided by NASA through Hubble Fellowship grant \#HF-01189.01
awarded by the Space Telescope Science Institute, which is operated by the Association
of Universities for Research in Astronomy. This publication makes use of data products from
the Two Micron All Sky Survey, which is a joint project of the University of Massachusetts and Infrared
Processing and Analysis Center/California Institute of Technology, funded by the National Aeronautics
and Space Administration and the National Science Foundation; the NASA/IPAC Infrared Science Archive, which
is operated by the Jet Propulsion Laboratory/California Institute of Technology, under contract with the 
National Aeronautics and Space Administration.
This research has made use of the ALADIN, VIZIER, SIMBAD databases, 
operated at CDS, Strasbourg, France.

\clearpage

\begin{table}
\caption{THE SAMPLE OF 8 LATE-M DWARF STARS 
  \label{sample}
  }
 $$
\begin{tabular}{llllllll} 
 \hline
 \hline
Name & $\alpha$$^{a}$  & $\delta$$^{a}$ &  SpT  &  $d$ &  $F_{\nu ,R}$ & $v\sin i$       &   $H\alpha$~EW    \\ 
     &           &          &       & (pc) &  ($\mu$Jy)    & (km~s$^{-1}$) &   (\AA)              \\ 
\hline 
LP 349-25AB  & 00 27 55.93 & $+$22 19 32.8 & M8.0$^{b}$ & 15   & 365$\pm$16     &   ~...    &   ~...     \\
2M 0109$+$29 & 01 09 21.70 & $+$29 49 25.6 & M9.5 & 19   & ~\,$<$54       &   ~\,7    &   ~\,0.6   \\
2M 0140$+$27 & 01 40 02.64 & $+$27 01 50.6 & M8.5 & 19   &  ~\,$<$20      &   ~\,6.5  &   ~\,3.0   \\
2M 0149$+$29 & 01 49 08.96 & $+$29 56 13.2 & M9.5 & 17   &  ~\,$<$140     &   12      &   ~\,5.5   \\
2M 0810$+$14 & 08 10 58.65 & $+$14 20 39.1 & M9.0 & 20   &  ~\,$<$39      &   11      &  12.3      \\
2M 1421$+$18 & 14 21 31.45 & $+$18 27 40.8 & M9.5 & 20   &  ~\,$<$42      &   ~...    &   ~\,3.6   \\
2M 1627$+$81 & 16 27 27.94 & $+$81 05 07.6 & M9.0 & 21   &  ~\,$<$60      &   ~...    &   ~\,6.1   \\
2M 1707$+$64 & 17 07 18.31 & $+$64 39 33.1 & M9.0 & 17   &  ~\,$<$60      &   ~...    &   ~\,9.8   \\
 \hline
\end{tabular}
 $$
  \begin{list}{}{}
  \item[]
$^{a}$ Right ascension and declination from the 2MASS catalogue \\
$^{b}$ LP 349-25A: M7.5/M8.0; LP 349-25B: M8.5/M9.0 \\
NOTE.-- References for source properties: \citet{forveille05, gizis00, gatewood, reid02};

  \end{list}
\end{table}

\clearpage

\end{document}